\documentclass[prd,aps,preprint,epsfig,floats]{revtex4}
\usepackage{graphics}
\usepackage{epsfig}

\usepackage{graphicx}
\def\Jo#1#2#3#4{{\it #1} {\bf #2}, #3 (#4)}

\def\circa#1{\,\raise.3ex\hbox{$#1$\kern-.75em\lower1ex\hbox{$\sim$}}\,}

\def\NPB{{\rm Nucl. Phys.} {\bf B}}

\def\PLB{{\rm Phys. Lett.}  {\bf B}}

\def\PRD{{\rm Phys. Rev.} {\bf D}}

\begin{document}

\preprint{\vbox{ \hbox{UMD-PP-06-053} }}
\title{\Large\bf Connecting Leptogenesis to CP Violation in Neutrino
Mixings in a Tri-bimaximal Mixing model }
\author{\bf R.N. Mohapatra and Hai-Bo Yu }

\affiliation{ Department of Physics, University of Maryland,
College Park, MD 20742, USA}

\date{September, 2006}

\begin{abstract} We show that in a
recently proposed $S_3$ model for tri-bimaximal mixing pattern for
neutrinos, CP violating phases in neutrino mixings are directly
responsible for lepton asymmetry $\epsilon_\ell$. In the exact
tri-bimaximal limit, $\epsilon_\ell$ is proportional to one of the
Majorana phases whereas in the presence of small deviations from
tri-bimaximal pattern, there are two contributions, one being
proportional to the Dirac phase and the other to one of the two
Majorana phases. In the second case, $\theta_{13}$ is nonzero and
correlated with the deviation from maximal atmospheric mixing.
 \end{abstract}
\maketitle
\section{Introduction}
Seesaw mechanism for understanding small neutrino
masses\cite{seesaw1} provides an interesting way to understand the
origin of matter-anti-matter asymmetry\cite{fuku} via the CP
violating decay of the heavy right-handed neutrinos combined with
B+L violation by electroweak sphalerons\cite{shap}. This raises the
very exciting possibility that better understanding of neutrino
masses and mixings may help to resolve one of the deepest mysteries
of the Universe i.e. the origin of matter. A lot of attention has
therefore been rightly focussed on trying to connect various ways of
understanding neutrino masses with leptogenesis and obtaining
constraints on seesaw scale physics, lightest neutrino masses
etc.\cite{lepto}. A very interesting question in this connection is
whether CP violating phases in neutrino mixings that can be probed
in long baseline as well as in neutrinoless double beta decay
experiments are the ones that are responsible for the
matter-anti-matter asymmetry. It turns out that in generic seesaw
models there is no apriori connection between them and it is
hoped that in a true theory of neutrino masses and mixings, such a
connection may exist.

 Attempts to find such models have been made in the
past\cite{endoh} but they usually require additional assumptions
about parameters not directly related to observations to establish a
direct connection between leptogenesis phase and low energy neutrino
phase. We repeat that by a direct connection, we mean the phase
responsible for lepton asymmetry of the Universe is the same one
that appears as either a Dirac or one of the two Majorana phases in
neutrino mixings. The nontriviality of this problem stems from two
facts: (i) in generic seesaw models, lepton asymmetry $\epsilon_\ell$
depends only a subset of the phases of Dirac mass
matrix $M_D$ whereas low energy phases in the neutrino mass matrix
involves all of them; and (ii)  the seesaw formula  ``scrambles''
 up the phases due to multiplication of matrices so that any direct
connection between low and high energy phases, if they exist at all becomes
difficult to discern.

 In this letter, we show that in a recently proposed
$S_3$ model\cite{yu} for tri-bimaximal neutrino mixing\cite{tbm},
the structure of the neutrino mass matrix is so constrained by  
symmetry that a direct connection between the leptogenesis phase 
and neutrino mixing phases emerges. Thus within the context of this 
model, a
measurement of the neutrino CP phases would provide a direct
understanding of the origin of matter. This appears to us to be an
interesting result. A future direction of work would
be to unify quarks into the model so that one may perhaps understand the
origin of quark CP violation as well.

The motivation for our work is the recent indication that present
neutrino oscillation data points to a leptonic mixing pattern given
by the PMNS matrix in the so-called tri-bimaximal form\cite{tbm}:
\begin{eqnarray}
U_{TB}~=~\pmatrix{\sqrt{\frac{2}{3}} & \frac{1}{\sqrt{3}} & 0\cr
-\frac{1}{\sqrt{6}} & \frac{1}{\sqrt{3}} & \frac{1}{\sqrt{2}}\cr
-\frac{1}{\sqrt{6}}& \frac{1}{\sqrt{3}} & -\frac{1}{\sqrt{2}}}.
\end{eqnarray}
This form is very suggestive of an underlying symmetry of leptons.
The true nature of the
symmetry is however far from clear, although there
are many interesting suggestions\cite{a4}\cite{o3}\cite{s3}. Our
interest here is in an $S_3$ model proposed in\cite{yu} where the key
flavor symmetry leading to tri-bimaximal mixing is the permutation
symmetry of three leptonic families. The resulting neutrino mass
matrix is characterized  by only three complex
parameters, whose absolute values are constrained by already existing
observations. We find that (i) in the exact tri-bimaximal
limit, when there is no Dirac phase, one of the two Majorana phases is
directly responsible for the lepton
asymmetry of the Universe; (ii) even after we include small departures
from the tri-bimaximal limit, the direct connection remains -- there are
then two contributions to $\epsilon_\ell$, one being proportional to the Dirac
phase and the other to one of the two Majorana phases. This direct
connection is possible due to the simple form of $M_D$ dictated by the
$S_3$ symmetry of the model and the assumptions that in case (i)
only one and in case (ii) only two right handed neutrinos dominate the seesaw
formula as well as the fact there is an $S_3$ symmetric type II
contribution to the neutrino masses in both cases. We elaborate on these
points below.

 This paper is organized as follows: in sec. 2, we review the
 salient features of the $S_3$ model of Ref.\cite{yu} for
 tri-bimaximal mixing; in section 3, we present a general discussion of 
leptogenesis in our model; in sec. 4, we calculate the baryon 
asymmetry in the exact tri-bimaximal mixing and establish the direct 
connection between one of the Majorana phases in the neutrino mixing and 
$\epsilon_\ell$; in sec. 5, we do the same for the case where we include
 deviations from tri-bimaximal limit and show the connection of 
$\epsilon_\ell$ to the Dirac and the Majorana phases; we summarize our 
results in sec. 6.

\section{The $S_3$ model}
We start with the Majorana neutrino mass matrix whose
diagonalization at the seesaw scale leads to the tri-bimaximal
mixing matrix:
\begin{eqnarray}
{\cal M}_\nu=\pmatrix{a' & b' & b'\cr b' & a'-c' & b'+c'\cr b' &
b'+c' & a'-c'}
\end{eqnarray}
where the elements are chosen to be complex. Diagonalizing this
matrix leads to the $U_{PMNS}$ of Eq. (1) and the neutrino masses:
$m_1=a'-b'; m_2=a'+2b'$ and $m_3=a'-b'-2c'$. Clearly if $|a'|\simeq
|b'| \ll |c'|$, we get a normal hierarchy for masses.
It was pointed out in Ref.\cite{yu} that the above
Majorana neutrino mass matrix can be realized in a combined type I
type II seesaw model with soft-broken $S_3$ family symmetry for
leptons. The type II contribution comes from an $S_3$ invariant
coupling $f_{\alpha\beta}L_\alpha L_\beta\Delta$,
\begin{eqnarray}
f=\pmatrix{f_a & f_b & f_b\cr f_b & f_a & f_b\cr f_b & f_b & f_a}
\end{eqnarray}
After the triplet Higgs field $\Delta$ gets vev and decouples, its
contribution to the light neutrino mass can written as
\begin{eqnarray}\label{mII}
M_{II}=\pmatrix{a' & b' & b'\cr b' & a' & b'\cr b' & b' & a'}
\end{eqnarray}
where $a'=\frac{v^2\sin^2\beta\lambda}{M_T}f_a$ and
$b'=\frac{v^2\sin^2\beta\lambda}{M_T}f_b$. We denote $M_T$ as the
mass of the triplet Higgs and $\lambda$ as the coupling constant
between the triplet and doublets in the superpotential.

Coming to the type I contribution, the Dirac mass matrix for
neutrinos comes from an $S_3$ invariant Yukawa coupling of the
form:
\begin{eqnarray}
{\cal L}_D~=~h_\nu[\overline{\nu_{R1}}H(L_e-L_\mu)+
\overline{\nu_{R2}}H(L_\mu-L_\tau)+\overline{\nu_{R3}}H(L_\tau-L_e)]+h.c.
\end{eqnarray} leading to
\begin{eqnarray}
Y_\nu=\pmatrix{h&-h&0\cr 0&h&-h\cr -h&0&h}.
\end{eqnarray}
In the limit of $|M_ {R1,R3}|\gg |M_{R2}|$, where a single
right-handed neutrino dominates the type I contribution, the mixed
type I+II seesaw formula
\begin{eqnarray}
{\cal M}_\nu~=~M_{II}-M^T_DM_{\nu R}^{-1}M_D,
\end{eqnarray}
 gives rise to the desired form for the neutrino Majorana mass matrix
which leads to the tri-bimaximal mixing\cite{yu}.

We can now do the phase counting in the model. When two of the above
right-handed neutrinos decouple, there is only one Yukawa coupling.
We can first redefine the phase of $\nu_{R2}$ so that its mass is
real and we then redefine all the lepton doublets by a common phase
which now makes the Dirac Yukawa coupling $h$ real.
 One cannot then do any more phase
redefinitions and we are left with two phases in the neutrino mass
matrix which in this basis reside in the entries $a'$ and $b'$ in
Eq.({\ref{mII}}). These two phases will appear as the Majorana phases
in the low energy mass matrix as we show below.

As far as the charged lepton masses are concerned, the symmetry needs to
be extended to $S_3\times (Z_2)^3$ to have a simple diagonal mass matrix
and  all their masses can be made real by
separate independent phase redefinition of the right-handed charged
leptons. No new phases enter the PMNS matrix. It turns
out that the $S_3\times (Z_2)^3$ symmetric version can also be derived
from an $S_4\times Z_2$ symmetry\cite{hlm} and this also does not effect
our phase counting.

Turning to the case where two of the right-handed neutrinos contribute
to ${\cal M}_\nu$, there are three phases in the light
neutrino mass matrix. This is because in this case there are two apriori
complex right-handed neutrino masses and only one of them together
with $h$ can be made real by phase redefinition as in the first case.
This leaves the phases of $a'$ and $b'$ and that of the second right
handed neutrino giving a total of three phases.
This case represents a deviation from the tri-bimaximal mixing with the
deviation being proportional to $|M_{R2}|/|M_{R3}|$. We will show in
sec. 4 that the new phase in this case appears as the Dirac phase.
Let us now proceed to discuss leptogenesis in both these cases.
As noted, we choose $f_a,f_b,M_{R3}$ to be complex
and $h,M_{R2}$ to be real, and express them as
 $f_a=|f_a|e^{i\phi_a},~f_b=|f_b|e^{i\phi_b},M_{R3}=M_3e^{-i\phi_3}$ and
$M_{R2}=M_2$.

\section{Leptogenesis in the $S_3$ model}
In this section, we present the calculation of lepton asymmetry in our
model and show that for the parameter range of interest from neutrino
mixing physics, one can explain the baryon asymmetry of the universe
whose present value is given by the WMAP observations\cite{spergel} to be
\begin{eqnarray}
\frac{n_B}{n_\gamma}=6.1{\pm0.2}\times 10^{-10}.
\end{eqnarray}

Let us start by reminding ourselves of some well known facts about
leptogenesis.
In the type I seesaw scenario, lepton asymmetry is generated by the
out-of-equilibrium decay of the right-handed neutrinos which participate
in the seesaw mechanism to give neutrino masses and mixings. Most of the
discussion of leptogenesis uses type I seesaw and there have been many
papers\cite{lepto} which have studied its  connection to neutrino masses
and mixings. In models with both type I\cite{seesaw1} and type II 
seesaw\cite{seesaw2} (induced by Higgs triplets\cite{valle}), the
presence of the triplet Higgs may also contribute to the lepton
asymmetry in two ways: either the decay of one or more
triplets\cite{ma} or the decay of right-handed neutrino with
triplets running in the
loop\cite{goran}\cite{king}. Our model involves both type
I and type II seesaw; however, it turns
out that the first contribution (i.e. the one from triplet decay) is
highly suppressed and only the lightest right-handed neutrino(sneutrino)
decay is important, which we compute below.

The asymmetry from the decay of the right-handed neutrino $\nu_{Ri}$
into a lepton(slepton) and a Higgs(Higgsino) is given by:
\begin{eqnarray}
\varepsilon_i=\frac{\Gamma[\nu_{Ri}\rightarrow lH(\tilde l\tilde
H)]-\Gamma[\nu_{Ri}\rightarrow  \bar  lH^*(\tilde{\bar l}\tilde
H^*)]}{\Gamma[\nu_{Ri}\rightarrow lH(\tilde l\tilde
H)]+\Gamma[\nu_{Ri}\rightarrow \bar lH^*(\tilde{\bar l}\tilde
H^*)]},
\end{eqnarray}
and  we also have the sneutrino $\tilde\nu_{Ri}$ decay
asymmetry, which we denote as $\tilde\varepsilon_i$. If one ignores the
supersymmetry breaking effects, one has
$\varepsilon_{i}=\tilde\varepsilon_{i}$.

In the basis where right-handed neutrinos mass matrix is diagonal,
the decay asymmetry of right-handed neutrino from type I
contribution is given by\cite{vissani}
\begin{eqnarray}\label{epsilonI}
\varepsilon^I_i=-\frac{1}{8\pi}\frac{1}{[Y'_{\nu}
Y'^\dag_\nu]_{ii}}\sum_j{\rm Im{[Y'_\nu
Y'^\dag_\nu]_{ij}^2}}F(\frac{M_j^2}{M_i^2}),
\end{eqnarray}

where $F(x)=\sqrt x(\frac{2}{x-1}+\ln[\frac{1+x}{x}])$ and for
$x\gg 1$, $F(x)\simeq\frac{3}{\sqrt x}$.

The type II contribution has been calculated and is given in Ref.
\cite{goran}\cite{king} to be
\begin{eqnarray}\label{epsilontype2}
\varepsilon^{II}_i=\frac{3}{8\pi}\frac{{\rm Im}[Y'_\nu
f^*Y'^T_\nu\mu]_{ii}}{[Y'_\nu
Y'^\dag_\nu]_{ii}M_i}\ln(1+\frac{M_i^2}{M_T^2}),
\end{eqnarray}
where $\mu\equiv\lambda M_T$ and $\lambda$ is the coupling between
triplet and two doublets in the superpotential. In general $\lambda$ is 
complex, but its
phase can be absorbed by rescaling phases of every elements of
matrix $f$ with same amount. We will treat it real in our
discussion.

The total contribution to the lepton asymmetry then becomes
\begin{eqnarray}
\varepsilon_i=\varepsilon^{I}_i+\varepsilon^{II}_i.
\end{eqnarray}
In our model, the lightest right-handed neutrino is $\nu_{R2}$, and
we will take $i=2$.

The generated $B-L$ asymmetry can be written as
\begin{eqnarray}
Y_{B-L}\equiv\frac{n_{B-L}}{s}
=-\eta(\varepsilon_2Y^{EQ}_{\nu_{R2}}+
\tilde{\varepsilon}_2Y^{EQ}_{\tilde{\nu}_{R2}}) \end{eqnarray}
where
\begin{eqnarray}
\nonumber Y^{EQ}_{\nu_{R2}}=\frac{n^{EQ}_{\nu_{R2}}}{s}
=\frac{3}{4}\frac{45\zeta(3)}{\pi^4g_{*s}}\\
Y^{EQ}_{\tilde{\nu}_{R2}}=\frac{n^{EQ}_{\tilde{\nu}_{R2}}}{s}
=\frac{45\zeta(3)}{\pi^4g_{*s}},
\end{eqnarray}
$g_{*s}$ is the effective degree of freedom contributing to entropy
$s$ with value $228.75$ in MSSM, and $\eta$ is the efficiency factor
for leptogenesis. Ignoring the SUSY breaking effect, we have
$\varepsilon_2=\tilde{\varepsilon}_2$ and $Y_{B-L}$ can be
simplified as
\begin{eqnarray}
Y_{B-L}=-\frac{7}{4}\frac{45\zeta(3)}{\pi^4g_{*s}}\eta\varepsilon_2.
\end{eqnarray}

Lepton number asymmetry produced by decay of right-handed
neutrino(sneutrino) can be converted to baryon number asymmetry by
sphaleron effect. The baryon number is related to the $B-L$
asymmetry $Y_{B-L}$ via
\begin{eqnarray}
Y_B=wY_{B-L},
\end{eqnarray}
where $w=\frac{8N_F+4N_H}{22N_F+13N_H}$ with $N_F$ as generations
of fermions and $N_H$ as the number of the Higgs doublet. In MSSM,
$N_F=3$ and $N_H=2$, one has $w=\frac{8}{23}$. Putting all this together,
we get the baryon to photon ratio to be
\begin{eqnarray}\label{nb/ngamma}
\frac{n_B}{n_\gamma}\simeq7.04Y_B=-1.04\times10^{-2}\varepsilon_2\eta.
\end{eqnarray}

The efficiency factor $\eta$ can be calculated by solving a set of
coupled Boltzmann equations(See for example
Refs.\cite{buchmuller}\cite{riotto}). We assume that to a good
approximation the efficiency factor depends only on a mass parameter
usually called the effective mass and the initial abundance of the
right-handed neutrino(sneutrino). We also use the result for $\eta$
in type I seesaw scenario. In our model, the effective mass for both
the cases discussed below, is given by
\begin{eqnarray}
\tilde m_2=\frac{[Y_\nu
Y^\dag_\nu]_{22}v^2\sin^2\beta}{M_2}=\frac{2h^2v^2\sin^2\beta}{M_2}\simeq
\sqrt {\Delta m^2_A}\simeq0.05{\rm eV},
\end{eqnarray}
which is larger than the equilibrium neutrino mass
$m_*=\frac{16\pi^{5/2}\sqrt
g_{*}}{3\sqrt5}\frac{v^2\sin^2\beta}{M_{pl}}\simeq1.50\times10^{-3}{\rm
eV}$, so it is in the strong washout region. In this region, the
dependence of efficiency factor on the initial abundance of
right-handed neutrino(senutrino) is small\cite{bdp}\cite{riotto}. We
take the approximation formula from Ref.\cite{riotto} to estimate
the efficiency factor for our model
\begin{eqnarray}
\frac{1}{\eta}\simeq\frac{3.3\times10^{-3}{\rm eV}}{\tilde
m_2}+(\frac{\tilde m_2}{0.55\times10^{-3}{\rm eV}})^{1.16},
\end{eqnarray}
and find $\eta\simeq 5.3\times 10^{-3}$, which we will use in the  
calculation of baryon to photon ratio for our model.

\section{Exact tri-bimaximal limit}
In this section, we establish the connection between $\epsilon_\ell$ and 
the low energy phase in the neutrino mixing.
In the limit of $|M_{R1,R3}|\rightarrow\infty$, light neutrino mass
matrix has the form that leads to tri-bimaximal mixing pattern. In
this limit, the contributions to lepton asymmetry from the
exchange of $\nu_{R1}$ and $\nu_{R3}$ in the loops 
are negligible. As far as neutrino masses go, $\nu_{R2}$
contribution dominates $\Delta m_A^2$ and triplet Higgs has the full
contribution to $\Delta m_\odot^2$. The observed values require that
$M_T\sim(10^1-10^2)M_2$. This triplet can go into loop of the decay
of $\nu_{R2}$ and its interference with tree level diagram of
$\nu_{R2}$ decay can generate lepton asymmetry.
In this case, Eq.(\ref{epsilontype2}) is simplified as
\begin{eqnarray}
\varepsilon^{II}_2=\frac{3}{8\pi}\frac{{\rm Im}[Y_\nu
f^*Y_\nu^T]_{22}\mu}{[Y_\nu
Y^\dag_\nu]_{22}M_2}\ln(1+\frac{M_2^2}{M_T^2}).
\end{eqnarray}
From Yukawa coupling matrices, one easily gets
\begin{eqnarray}
{\rm Im} [Y_\nu
f^*Y_\nu^T]_{22}=2h^2(|f_b|\sin\phi_b-|f_a|\sin\phi_a)
\end{eqnarray}
\begin{eqnarray}
[Y_\nu Y_\nu^\dag]_{22}=2h^2.
\end{eqnarray}
We also have
\begin{eqnarray}
|f_a|=a\frac{M_T}{v^2\sin^2\beta\lambda},
|f_b|=b\frac{M_T}{v^2\sin^2\beta\lambda}
\end{eqnarray}
where $a\equiv|a'|$ and $b\equiv|b'|$, and $\varepsilon^{II}_2$
can be written as
\begin{eqnarray}
\varepsilon^{II}_2=\frac{3}{8\pi}
\frac{(b\sin\phi_b-a\sin\phi_a)M_2}{v^2\sin^2\beta}
\frac{M_T^2}{M_2^2}\ln(1+\frac{M_2^2}{M_T^2}).
\end{eqnarray}
Note that in the tri-bimaximal limit,
\begin{eqnarray}
M_{\nu}=\pmatrix{a e^{i\phi_a} & b e^{i\phi_b} & b e^{i\phi_b}\cr b
e^{i\phi_b}& a e^{i\phi_a}-c & b e^{i\phi_b}+c\cr b e^{i\phi_b} & b
e^{i\phi_b}+c & a e^{i\phi_a}-c},
\end{eqnarray}
which can be diagnolized by $U_{TB}$
\begin{eqnarray}
U_{TB}^TM_{\nu}U_{TB}=\pmatrix{ae^{i\phi_a}-be^{i\phi_b}&0&0\cr
0&ae^{i\phi_a}+2be^{i\phi_b}&0\cr
0&0&-2c+ae^{i\phi_a}-be^{i\phi_b}}.
\end{eqnarray}
Therefore one of the Majorana phases is given by
\begin{eqnarray}
\varphi_1\simeq{\rm Arc}\sin[\frac{a\sin\phi_a-b\sin\phi_b}{m_1}]
\end{eqnarray}
up to $O(\sqrt{\frac{\Delta m_{\odot}^2}{\Delta m_A^2}})$. And for
$M_T\geq (10^1-10^2)M_2$, one has
$\frac{M_T^2}{M_2^2}\ln(1+\frac{M_2^2}{M_T^2})\simeq1$. So the
lepton asymmetry can be written as
\begin{eqnarray}\label{epsilonII}
\varepsilon^{II}_2\simeq-\frac{3}{8\pi}
\frac{m_1M_2\sin\varphi_1}{v^2\sin^2\beta}.
\end{eqnarray}
Thus we see that the Majorana phase $\varphi_1$ directly gives the lepton
asymmetry, as noted in the introduction. This is the first main result of
this paper.

To estimate the value of the baryon to photon ratio, we note that
in this case $\varepsilon^{I}_2\simeq0$ and
$\varepsilon_2=\varepsilon^{II}_2$, using Eq.(\ref{nb/ngamma}) and
Eq.(\ref{epsilonII}), giving
\begin{eqnarray}
\frac{n_B}{n_\gamma}\simeq
6.1\times10^{-10}(\frac{m_1}{2.8\times10^{-3}{\rm
eV}})(\frac{M_2}{10^{12}{\rm
GeV}})(\frac{\sin\varphi_1}{1})(\frac{\eta}{5\times10^{-3}}),
\end{eqnarray}
where we take $v=170{\rm Gev}$ and $\tan\beta=10$.
To get the right range for baryon to photon ratio, the
lightest right-handed neutrino mass should be larger than about
$10^{12}{\rm GeV}$. Strict lower bound is on the product $m_1M_2\geq 2.8$ 
GeV$^2$. The thermal production of $\nu_{R2}$ requires a
reheat temperature of the Universe
after inflation be $T_{reh}\gtrsim10^{12}-10^{13}{\rm GeV}$.

If we take as upper bound on $M_2$ to be $10^{14}{\rm GeV}$ required to 
fit the atmospheric neutrino data, to get
right baryon to photon ratio, we have to have a lower bound of
$m_1\sim 10^{-5}{\rm eV}$. On the other hand, if we take
$M_2\sim10^{14}{\rm GeV}$ and $m_1\sim 10^{-3}{\rm eV}$, we get the
lower bound of $\sin\varphi_1$ as $\sim 10^{-2}$.

\section{Departure from tri-bimaximal mixing and new contribution to
leptogenesis}

In this section, we consider the case when we relax the mass
constraint on the right-handed neutrinos and assume that
$|M_{R2}|<|M_{R3}|\ll|M_{R1}|$. This will lead to departures from
the exact tri-bimaximal mixing pattern\cite{werner}. In this case,
there are three independent phases as noted above.

While the type II contribution to neutrino mass matrix in this case
remains
the same as in the exact tri-bimaximal case, the type I contribution
changes and is given by
\begin{eqnarray}
M_I=-M^T_DM_{\nu R}^{-1}M_D=-\pmatrix{\sigma e^{i\phi_3}&0&-\sigma
e^{i\phi_3} \cr 0&c&-c\cr -\sigma e^{i\phi_3}&-c&c+\sigma
e^{i\phi_3}},
\end{eqnarray}
where $c\equiv\frac{h^2}{M_2}v^2\sin^2\beta$ and
$\sigma\equiv\frac{h^2}{M_3}v^2\sin^2\beta$.

Combining the contributions from type I and type II, the light
neutrino mass matrix is found to be
\begin{eqnarray}
M_{\nu}=\pmatrix{a e^{i\phi_a}-\sigma e^{i\phi_3} & b e^{i\phi_b}
& b e^{i\phi_b}+\sigma e^{i\phi_3}\cr b e^{i\phi_b}& a
e^{i\phi_a}-c & b e^{i\phi_b}+c\cr b e^{i\phi_b}+\sigma
e^{i\phi_3} & b e^{i\phi_b}+c & a e^{i\phi_a}-c-\sigma
e^{i\phi_3}}.
\end{eqnarray}
To diagnolize $M_\nu$, we first consider $U^{\dag}_{TB}M_{\nu}^\dag
M_{\nu} U_{TB}$. The off-diagonal elements of
$U^{\dag}_{TB}M_{\nu}^\dag M_{\nu} U_{TB}$ are all zeros except
$1-3$ and $3-1$ entries,
\begin{eqnarray}
[U^{\dag}_{TB}M_{\nu}^\dag M_{\nu}
U_{TB}]_{13}=\sqrt3\sigma(ce^{-i\phi_3}+
\sigma-a\cos(\phi_3-\phi_a)+b\cos(\phi_3-\phi_b)).
\end{eqnarray}

To further diagnolize $U^{\dag}_{TB}M_{\nu}^\dag M_{\nu} U_{TB}$,
one needs another rotation in the $1-3$ plane. Because of the
normal hierarchical mass spectrum of the light neutrinos, one has
$c\gg a\simeq b$, and also $c\gg \sigma$ due to small upper bound
of $\sin\theta_{13}$ value.
In these approximation, the unitarity matrix in $1-3$
plane is
\begin{eqnarray}
V=\pmatrix{1&0&\xi\cr 0&1&0 \cr -\xi e^{i\phi_3}&0&e^{i\phi_3}}
\end{eqnarray}
where $\xi\simeq\frac{\sqrt3\sigma}{4c}$. Now the mixing matrix is
given by $U=U_{TB}V$,

\begin{eqnarray}\label{U}
U=\pmatrix{\sqrt{\frac{2}{3}}&\frac{1}{\sqrt
3}&\sqrt{\frac{2}{3}}\xi\cr
-\frac{1}{\sqrt6}-\frac{e^{i\phi_3}\xi}{\sqrt2}&
\frac{1}{\sqrt3}&\frac{e^{i\phi_3}}{\sqrt2}-\frac{\xi}{\sqrt6}\cr
-\frac{1}{\sqrt6}+\frac{e^{i\phi_3}}{\sqrt2}&
\frac{1}{\sqrt3}&-\frac{e^{i\phi_3}}{\sqrt2}-\frac{\xi}{\sqrt6}}.
\end{eqnarray}

From this mixing matrix, we can read
$\tan\theta_{12}=\frac{|U_{12}|}{|U_{11}|}=\frac{1}{\sqrt2}$,
$\sin\theta_{13}=\sqrt{\frac{2}{3}}\xi$ and
$\tan\theta_{23}=\frac{|U_{23}|}{|U_{33}|}\simeq
1-\frac{2\xi}{\sqrt3}\cos\phi_3$. Note the correlation between
$\theta_{13}$ and the
departure of $\theta_{23}$ from its maximal value.  For the Dirac
phase, we use the
Jarskog invariant\cite{jarlskog} to extract it from above mixing
matrix $J_{CP}={\rm
Im}[U_{11}U_{22}U^*_{12}U^*_{21}]=\frac{1}{8}\sin2\theta_{13}
\sin2\theta_{23}\cos\theta_{13}\sin\delta$.
From Eq.(\ref{U}), one can easily get
\begin{eqnarray}
{\rm
Im}[U_{11}U_{22}U^*_{12}U^*_{21}]&=&\frac{\xi}{3\sqrt3}\sin\phi_3\\
\frac{1}{8}\sin2\theta_{13}\sin2\theta_{23}\cos\theta_{13}\sin\delta
&=&\frac{\xi}{3\sqrt3}\sin\delta.
\end{eqnarray}
Therefore we have $\delta\simeq\phi_3$. Remarkably, although this
model has three independent CP phase at the seesaw scale, the low
energy scale Dirac phase is equal to one of the phases at the high
energy scale up to $O(\sqrt\frac{\Delta m_\odot^2}{\Delta m_A^2})$.
This is independent of the way to assign these three phases.

Coming to the calculation of lepton asymmetry in this case, with
$|M_{R2}|<|M_{R3}|\ll|M_{R1}|$ limit, besides the contribution
from type II to the lepton asymmetry, we should also consider the
contribution from type I. From Eq.(\ref{epsilonI}), we have
\begin{eqnarray}\label{epsilon2}
\varepsilon^I_2=-\frac{1}{8\pi}\frac{1}{[Y'_\nu
Y'^\dag_\nu]_{22}}{\rm Im}[Y'_\nu Y'^\dag_\nu]_{23}^2
F(\frac{M_3^2}{M_2^2}),
\end{eqnarray}
and $Y'_\nu=U_R^\dag Y_\nu$, where $U_R$ is to diagnolize the
right-handed neutrino mass matrix.

In the two light right-handed neutrinos limit, the phase of the mass
of the heaviest right-handed neutrino is irrelevant to the lepton
asymmetry and one can take $U_R={\rm diag}(1,1,e^{i\phi_3/2})$. 
Therefore we have $[Y'_\nu
Y'^\dag_\nu]_{23}=-h^2e^{i\phi_3/2},[Y'_\nu Y'^\dag_\nu]_{22}=2h^2$
and $F(\frac{M_3^2}{M_2^2})\simeq3\frac{M_2}{M_3}$, and plugging them
into Eq.({\ref{epsilon2}}), we get
\begin{eqnarray}
\varepsilon^I_2=-\frac{3}{8\pi}\frac{h^2}{2}\sin\phi_3\frac{M_2}{M_3}
\end{eqnarray}
Notice that $\delta\simeq\phi_3$,
$\sin\theta_{13}=\sqrt{\frac{2}{3}}\xi=\frac{\sqrt2}{4}\frac{M_2}{M_3}$
, $\Delta m_A^2\simeq 4c^2$ and $c=\frac{h^2}{M_2}v^2\sin^2\beta$,
one can rewrite $\varepsilon^I_2$ as function of the low energy
scale observables,
\begin{eqnarray}
\varepsilon^I_2\simeq-\frac{3}{8\pi}\frac{\sqrt{\Delta
m_A^2}M_2}{\sqrt{2}v^2\sin^2\beta}\sin\delta\sin\theta_{13}.
\end{eqnarray}

Combining the contribution from $\varepsilon_2^{II}$ given in
Eq.(\ref{epsilonII}), we have
\begin{eqnarray}
\varepsilon_2=\varepsilon^{II}_2+\varepsilon^I_2\simeq-
\frac{3}{8\pi}\frac{M_2}{v^2\sin^2\beta}[\sqrt{\frac{\Delta
m_A^2}{2}}\sin\delta\sin\theta_{13}+m_1\sin\varphi_1]
\end{eqnarray}
We again see that the phases in the leptogenesis formula are the
same phases in the neutrino mixing matrix- one Dirac and one
Majorana. This is the second main result of our paper. In this case
also one can get the right value for the baryon to photon ratio by
choosing the $M_2$ masses.

\section{Conclusion} In conclusion, we have shown that in a model
for tri-bimaximal neutrino mixing derived from an $S_3$ permutation
symmetry among lepton generations, the observable neutrino phases at
low energies are directly responsible for the origin of matter (up to 
 small corrections of order $\sqrt{\frac{\Delta m^2_\odot}{\Delta 
m^2_A}}$)Therefore, a
measurement of the low energy neutrino phase in this model will provide a 
direct
understanding of the high temperature early universe phenomenon of
the origin of matter. This model is especially interesting in view
of the fact that tri-bimaximal mixing pattern very closely resembles
current experimental observations. Measurement of $\theta_{13}$ and 
$\theta_{23}$ can provide test of the tri-bimaximal mixing. If this 
pattern gets confirmed,  experimental search for leptonic phases will 
become a matter of deep interest since it may hold the key to a 
fundamental mystery of cosmology.

 This work is supported by the National Science Foundation grant
no. PHY-0354401

\end{document}